\documentclass[aps,pra,twocolumn,showpacs,groupedaddress]{revtex4}
\usepackage{graphicx}
\begin{document}
\title{Raman coupler for a trapped two-component quantum-degenerate Fermi
gas}

\author{Sierk P\"otting$^{1,2,3}$\thanks{email: sierk.poetting@optics.arizona.edu},
        Marcus Cramer$^{1,4}$,
        Weiping Zhang$^{1}$,
        and Pierre Meystre$^{1}$}
\affiliation{ $^1$Optical Sciences Center, The University of
Arizona,
Tucson, AZ 85721 \\
$^2$Max--Planck--Institut f\"ur Quantenoptik, 85748 Garching,
Germany \\
$^3$Sektion Physik, Universit\"at M\"unchen, 80333 M\"unchen,
Germany \\
{$^4$Fachbereich Physik der Philipps--Universit\"at, 35032
Marburg, Germany}}
\date{\today}

\begin{abstract}
We investigate theoretically the Raman coupling between two
internal states of a trapped low-density quantum-degenerate Fermi
gas. In general, the trap frequencies associated with the two
internal states can be different, leading to the onset of
collapses and revivals in the population difference $\Delta N$ of
the two internal states. This behavior can be changed drastically
by two-body collisions. In particular, we show that under
appropriate conditions they can suppress the dephasing leading to
the collapse of $\Delta N$, and restore almost full Rabi
oscillations between the two internal states. These results are
compared and contrasted to those for a quantum-degenerate bosonic
gas.
\end{abstract}

\maketitle

%===========================================
\section{Introduction}
\label{sec:intro}
%===========================================

Quantum-degenerate samples of low-density fermionic atomic gases
\cite{Fermi,BCS} are arguably even more interesting than their
bosonic counterparts \cite{BEC}, due to the fundamental role
played by the Pauli exclusion principle in their dynamics. This
principle renders their experimental realization particularly
difficult, since in the simplest case, the collisions that are
essential in evaporative cooling largely disappear as the
temperature of the sample goes to zero. For this reason, more
elaborate techniques, involving e.g. the use of several isotopes,
or sympathetic cooling via a bosonic system, have been used to
achieve degeneracy \cite{Fermi}. Collisions also leave behind
holes in the Fermi sea \cite{holes}; these holes are difficult to
fill and are believed to limit the temperatures that can be
achieved to about 0.2 $T_F$, where $T_F$ is the Fermi
temperature.

From a theoretical viewpoint, Fermi systems also present a number
of difficult challenges. In particular, they are not amenable to a
mean-field description. Hence they cannot be analyzed in a
classical-like formalism such as the Gross-Pitaevskii equation,
which has proven remarkably powerful in describing many aspects of
quantum-degenerate bosonic systems. On the other hand, the
additional complexity of Fermi systems also offers much promise.
One can hope to be able to manipulate them into strongly
non-classical states, with potential applications in atom
interferometry. Also, the nonlinear mixing of fermionic matter
waves is expected to be very different from the bosonic case. As
such, Fermi systems promise the extension of nonlinear atom
optics \cite{atomoptics} to a regime without counterpart in
traditional nonlinear optics.

Just as is the case for bosons, a cornerstone of the manipulation
of fermionic matter waves is their interaction with light. In this
paper, we discuss the specific situation where transitions between
two internal states of a trapped quantum-degenerate Fermi system
at zero temperature are induced by Raman coupling. A
straightforward generalization of this model could be used to
describe an output coupler for an atom laser \cite{atomlaser}. Our
goal is two-fold: first, to understand the difference between the
bosonic and fermionic dynamics; and second, to determine the role
of two-body collisions on the evolution of the system.

We first consider the dynamics of the system in the absence of
collisions. Section \ref{sec:model} introduces our model and
derives the Heisenberg equations of motion for the relevant atomic
fields, and section III compares the resulting dynamics with those
for a corresponding Bose gas. Collisions are introduced in section
IV. The resulting equations of motion are solved numerically in
the framework of a time-dependent Hartree-Fock theory for the case
of fermions, and in a standard mean-field theory for a bosonic
sample. Again, we give a detailed comparison of the two
situations, and illustrate how collisions can change the fermionic
dynamics in a non-trivial fashion. Finally, Section
\ref{sec:summary} is a summary and conclusion.

%===========================================
\section{Model}
\label{sec:model}
%===========================================

\begin{figure}
\begin{center}
\includegraphics[width=0.5\columnwidth]{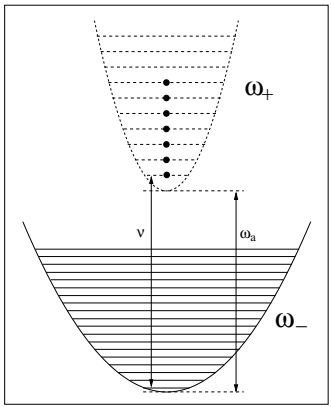}
\vspace{.3 cm} \caption{Two-component Fermi gas in a harmonic
trap. The trapping frequencies $\omega_+$ and $\omega_-$
correspond to the two internal states, which are coupled via a
spin-flip transition at frequency $\nu$ resonant with the
frequency difference of the trap ground states.
\label{fig:figure1} }
\end{center}
\end{figure}

We consider a two-component quantum-degenerate atomic system
trapped in a one-dimensional, harmonic potential with each
component corresponding, e.g., to one internal hyperfine spin
state. In general, the coupling of the atoms to the trapping field
is different for the two (spin) components $|+\rangle$ and
$|-\rangle$, so that they see trapping potentials of different
frequencies $\omega_+$ and $\omega_-$. The two internal states are
coupled by a Raman-type interaction of frequency $\nu$ equal to
the spin-flip transition frequency of the atoms in the ground
state of the two trapping potentials. This model, which is
summarized on the diagram of Fig. \ref{fig:figure1}, is described
by the second-quantized Hamiltonian

\begin{eqnarray}
\label{equ:Hamiltonian} {\hat {\cal H}} &=& \int {d}x\,
\hat{\Psi}^\dagger_+(x) H_+ \hat{\Psi}_+(x)
+ \int  {d}x\, \hat{\Psi}^\dagger_-(x) H_- \hat{\Psi}_-(x) \nonumber \\
&+& \hbar g \int  {d}x\, \left[ e^{-i\nu t}\hat{\Psi}^\dagger_+(x)
\hat{\Psi}_-(x) + h.c. \right],
\end{eqnarray}
where $\hbar g$ is the Raman coupling strength. The
first-quantized Hamiltonian describing the trapping potentials
associated with the internal states $|+\rangle$ and $|-\rangle$ is
\begin{equation}
H_\pm = -\frac{\hbar^2}{2m} \frac{\partial^2}{\partial x^2} +
\frac{1}{2} m \omega_\pm^2 x^2 + E_\pm,
\end{equation}
with $E_\pm$ being the energy of the internal state $|\pm\rangle$.
The Raman resonance condition is therefore, with
$\omega_a=E_+-E_-$,
\begin{equation}
\nu = \omega_a + (\omega_+ - \omega_-)/2 .
\end{equation}

We remark that this resonance condition effects only the coupling
between the two trap ground states: the coupling between all other
levels is off-resonant for $\omega_+ \neq \omega_-$. Hence,
introducing a small detuning even for the ground states does not
significantly alter the dynamics of the system.

The atomic field operators corresponding to the two traps obey the
fermionic, respectively bosonic (anti)commutation relations
\begin{eqnarray}
\label{equ:commutator}
\left[\hat{\Psi}_i(x),\hat{\Psi}_j^\dagger(x')\right]_\pm
& = & \delta_{ij}\delta(x-x'), \nonumber\\
\left[\hat{\Psi}_i(x),\hat{\Psi}_j(x')\right]_\pm & = & 0, \nonumber \\
\left[\hat{\Psi}_i^\dagger(x),\hat{\Psi}_j^\dagger(x')\right]_\pm
&=& 0,
\end{eqnarray}
where $i,j=\{+,-\}$.

For the harmonic potentials at hand, the Heisenberg equations of
motion for the atomic field operators take the same form,
independently of whether the atoms are bosonic or fermionic. It is
convenient to expand them in terms of eigenstates $\{u_n(x)\}$ of
one of the trap Hamiltonians  $H_\pm$, say, $H_+$ for
concreteness, as
\begin{eqnarray}
\label{equ:expansionsame}
\hat{\Psi}_+(x,t) &=& \sum_n u_n(x) \hat{a}_n(t), \nonumber \\
\hat{\Psi}_-(x,t) &=& \sum_n u_n(x) \hat{b}_n(t),
\end{eqnarray}
where the $\hat{a}_n$'s and $\hat{b}_n$'s satisfy either fermionic
or bosonic commutation relations. In both cases, this expansion
readily yields the Heisenberg equations of motion
\begin{eqnarray}
\label{equ:finalsame} i\frac{ d\hat{a}_n}{d \tau} &=& A_n
\hat{a}_n
 + \tilde{g} \hat{b}_n, \nonumber\\
i\frac{d \hat{b}_n}{d\tau} &=& B_n \hat{b}_n + C_n \hat{b}_{n+2} +
D_n \hat{b}_{n-2} + \tilde{g} \hat{a}_n,
\end{eqnarray}
where we have introduced the coefficients
\begin{eqnarray}
\label{equ:abbreviationlinear}
A_n &=& \frac{1}{2}\left(\beta-1\right)+n ,\nonumber\\
B_n &=& \frac{1}{4}\left(\beta^2-1\right)\left(2n+1\right)+n ,\nonumber\\
C_n &=& \frac{1}{4}\left(\beta^2-1\right)\sqrt{(n+2)(n+1)} ,\nonumber\\
D_n &=& \frac{1}{4}\left(\beta^2-1\right)\sqrt{n(n-1)},
\end{eqnarray}
and the ratio
\begin{equation}
\beta= \omega_+/\omega_- \end{equation} of the trap frequencies.
The dimensionless time $\tau$ is scaled to $\omega_+$, $\tau=
\omega_+ t$, and so is the dimensionless coupling strength
$\tilde{g} = g/\omega_+$.

We emphasize that while the operator $\hat{a}_n$ describes the
annihilation of atoms in level $n$ of the upper trap, a similar
interpretation of the $\hat{b}_n$'s is not possible, since they
result from the expansion of the field operator of atoms in the
internal state $|-\rangle$ on the basis of the ``+''-trap.
Denoting the eigenstates of the single-atom Hamiltonian of the
lower trap as $\{v_n(x)\}$, the ``true'' annihilation operators
${\hat c}_n$ associated with the trapped atoms in the $|-\rangle$
internal state are related to the $\hat{b}_n$'s by the mapping
\begin{eqnarray}
\label{equ:operator} \hat{c}_n(t)=\sum_n T_{nm} \hat{b}_m(t),
\end{eqnarray}
where the mapping matrix element $T_{nm}$ is the overlap integral
\begin{equation}
T_{nm}=\int {d}x\, v_n(x) u_m(x).
\end{equation}

%===========================================
\section{Dynamics}
\label{sec:dynamics}
%===========================================

In this section, we compare the dynamics of ideal noninteracting
bosonic and fermionic systems evolving under the influence of the
Raman coupling. We proceed by numerically solving the Heisenberg
equations of motion (\ref{equ:finalsame}) for a sample of $N$
atoms initially in the internal state $|+\rangle$ and at
temperature $T=0$. For bosonic atoms, all atoms are therefore
initially in the ``+''-trap ground state, while for fermions they
fill the lowest $N$ trap levels. The corresponding initial states
are correspondingly
\begin{equation}
\label{equ:initialstate} |\psi_F(0)\rangle =
\prod_{i=0}^{N-1}\hat{a}_i^\dagger|0\rangle_+ \otimes |0\rangle_-,
\end{equation}
in the case of fermions, and
\begin{equation}
\label{equ:initialstate1} |\psi_B(0)\rangle = \frac{1}{\sqrt{N!}}
\hat{a}_0^{\dagger N} |0 \rangle_+ \otimes |0\rangle_-
\end{equation}
for bosonic atoms.

We consider, first, the case of noninteracting fermionic atoms.
For trap frequencies approximately equal, $\beta \simeq 1$, Eq.
(\ref{equ:finalsame}) suggests the existence of two limiting
situations, at least in the case of fermions. (We will revisit
this point when discussing low-temperature bosonic systems.) In
the first one, which we call the ``strong-coupling regime'' in the
following, $\tilde{g} \approx N$, so that the inter-trap coupling
dominates the dynamics and the intra-trap coupling terms
$\hat{b}_{n\pm2}$ can largely be ignored. In contrast, the
``weak-coupling regime'' $\tilde{g} \ll N$ is dominated by
intra-trap coupling.

\begin{figure}
\begin{center}
\includegraphics[width=0.9\columnwidth]{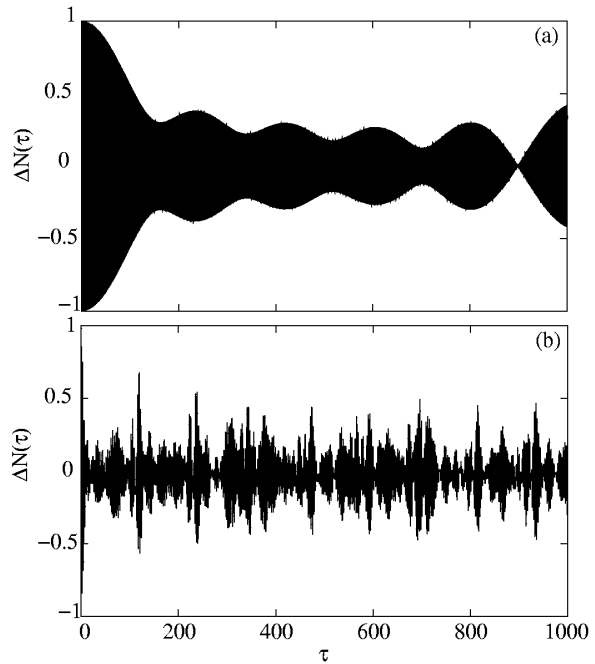}
\vspace{.3cm}
\caption{$\Delta N(\tau)$ for $N=10$ fermions and trap ratio
  $\beta=0.9$, as obtained from a numerical integration of
  Eqs. (\ref{equ:finalsame}): (a) strong coupling regime with
  $\tilde{g}=10.0$; (b) weak coupling regime with
   $\tilde{g}=1.0$. The dimensionless time $\tau$ is
  in units of $1/\omega_+$.
\label{fig:figure2} }
\end{center}
\end{figure}

As a first measure of the system dynamics, Fig. \ref{fig:figure2}
shows the difference
\begin{eqnarray}
\label{equ:signal} \Delta N (\tau) & = & \frac{1}{N} \int {d}x\,
\left[
  \left\langle\hat{\Psi}_+^\dagger(x) \hat{\Psi}_+(x)\right\rangle
 -\left\langle \hat{\Psi}_-^\dagger(x)\hat{\Psi}_-(x)\right\rangle
\right]
\nonumber \\
& = & \frac{1}{N} \sum_{n} \left(\left\langle
\hat{a}_n^\dagger(\tau) \hat{a}_n(\tau)\right\rangle -
\left\langle \hat{c}_n^\dagger(\tau)
\hat{c}_n(\tau)\right\rangle\right).
\end{eqnarray}
between the populations of the ``+'' and ``-'' traps. Fig.
\ref{fig:figure2}a is for the strong-coupling regime, and Fig.
\ref{fig:figure2}b for the weak-coupling regime.

One can gain some intuitive understanding of the strong-coupling
regime by remarking that in that regime, intra-trap transitions
remain small, so that the Raman coupling is predominantly between
levels of the two traps with equal quantum number $n$. To lowest
order, these transitions are all at the Rabi frequency $\tilde g$.
However, this simplest description cannot explain the result of
Fig. \ref{fig:figure2}a. Rather, it is necessary to include at least their
lowest-order corrections, i.e,
\begin{equation}
\label{equ:omegan} \Omega_n = \sqrt{{\tilde g}^2 + \frac{1}{4}\left(A_n -
B_n \right)^2} \simeq {\tilde g} + \left (\frac{(\beta-1)^2}{8
{\tilde g}}\right ) n^2.
\end{equation}
Such an $n$-dependence of Rabi frequencies is known to lead to
collapse and revival phenomena, as was first discussed in the
context of the Jaynes-Cummings model \cite{cummings}, where
$\Omega_n \propto {\sqrt n}$. This is precisely the type of
behavior exhibited by $\Delta N$ in the strong-coupling regime.
Because of the $n^2$-dependence of $\Omega_n$, it is expected that
the lowest trap levels, i.e. the atoms in the deep Fermi sea,
play a dominant role in the appearance of the revivals. We verified
that the populations of the lowest trap levels indeed oscillate
more or less in phase, while those of higher $n$ levels dephase rapidly.

We remark that both collapses and revivals of $\Delta N$ disappear
when the two trap frequencies are identical, since for $\beta=1$,
we have $A_n=B_n=n$ and hence $\Omega_n= {\tilde g}$. In addition,
intra-trap transitions vanish in that case, due to $C_n=D_n=0$.

Fig. \ref{fig:figure2}b shows the inversion $\Delta N$ between the
total trap populations in the weak-coupling regime, $\tilde{g} \ll
N$. In this limit, inter-trap and intra-trap coupling occur on
similar timescales. Immediately following a Raman transition from
the $|+\rangle$ to the $|-\rangle$ internal state, the population
of the ``-''-trap starts to undergo a redistribution between its
levels. The combined effects of the intra- and inter-trap
transitions result in that case in a random-looking evolution of
$\Delta N(t)$ of Fig. \ref{fig:figure2}b.

We now briefly turn to the case of a Bose gas. For a sample at
zero-temperature, $T=0$, and initially in the internal state
$|+\rangle$, all atoms are in the ground state of the ``+''-trap
at $\tau=0$. As a result, the strong-coupling regime is
characterized by almost perfect Rabi oscillations of the atomic
population between the two trap ground states, with a very small
fraction of the atoms coupling to higher modes due to intra-trap
transitions. This behavior is also largely preserved for $\beta^2
-1 \ll {\tilde g} \ll N$. (We recall that for fermions the
right-hand side of this inequality corresponds to the
weak-coupling regime, dominated by intra-trap transitions.) This
difference between bosons and fermions can readily understood from
Eqs. (\ref{equ:finalsame}), which show that in the $T=0$ bosonic
case, intra-trap coupling first occurs between the levels $n=0$
and $n=2$ of the ``-''-trap, with coupling coefficient $D_2
=(\beta^2-1)/4$. As long as this coupling remains small compared
to the inter-trap coupling ${\tilde g}$, the system acts
effectively as a two-mode system. In other words, for low
temperature bosonic systems, the weak-coupling regime is not
characterized by ${\tilde g} \approx N$ as is the case for
fermions, but rather by ${\tilde g}\ll \beta^2 -1$.

As is to be expected, the difference between fermionic and bosonic
systems is reduced as $T$ is increased. At first, the sharp edge
of the Fermi-Dirac distribution softens, resulting in slightly
reduced (strong-coupling) collapses and revivals of the fermionic
system. On the other hand, for $T \neq 0$, bosons occupy higher
trap states, resulting in a spread in Rabi frequencies
participating in the population difference signal. This in turn
leads to collapses and revivals rather than the perfect $T=0$ Rabi
oscillations. Increasing the temperature further leads of course
to undistinguishable behaviors of the bosonic and fermionic
systems.

%===========================================
\section{Collisions}
\label{sec:collision}
%===========================================

In this section, we discuss the effect of collisions on the
preceding results. Collisions are of course central to the
dynamics of quantum-degenerate atomic systems. They are essential
in the evaporative cooling of the sample, and also provide a
nonlinearity that can lead to the nonlinear mixing of matter
waves. In bosonic systems, much new physics can be studied, e.g.
by changing the sign of the scattering length of $s$-wave
collisions. In the case of fermions the creation of holes in the
Fermi sea, and the filling of these holes by additional collisions
results in a heating that appears to fundamentally limit the
temperatures at which these samples can be cooled \cite{holes}.
First, we discuss the way collisions impact the operation of the
Raman coupler in the case of fermions, and later compare these
results with those for a bosonic sample.

It is well known that in fermionic atoms, the Pauli exclusion
principle forbids the existence of $s$-wave scattering between
atoms in the same internal state. In addition, $p$-wave scattering
is generally negligible. Hence two-body collisions are described
by the Hamiltonian
\begin{equation}
\label{equ:hcolferm} {\hat {\cal H}}_{\rm {col}} = U_0 \int {d}x\,
\hat{\Psi}^\dagger_+(x)\hat{\Psi}^\dagger_-(x)
\hat{\Psi}_-(x)\hat{\Psi}_+(x),
\end{equation}
where $U_0=4\pi\hbar^2a\rho/m$ is the interaction strength with
$a$ being the s-wave scattering length and $\rho$ the
characteristic density of the system. Again, we expand the field
operators according to Eq. (\ref{equ:expansionsame}) in terms of
the basis $\{u_n\}$ and obtain
\begin{equation}
{\hat H}_{\rm{col}} = U_0 \sum_{i,j,k,l} U_{ijkl}\mbox{ }
\hat{a}_i^\dagger \hat{b}_j^\dagger \hat{b}_k \hat{a}_l,
\end{equation}
where the matrix element
\begin{equation}
\label{equ:scatmat}
U_{ijkl} = \int {d}x\, u_i(x) u_j(x) u_k(x) u_l(x)
\end{equation}
characterizes the scattering between different levels. We note
that $U_{ijkl}$ is symmetric under permutations.

In the presence of this quartic Hamiltonian, the Heisenberg
equations of motion for the operators $a_n$ and $b_n$ involve
cubic combinations of operators. To close this system of
equations, we invoke a time-dependent Hartree-Fock ansatz,
which has proved to be very successful in the treatment of 
many-particle quantum systems \cite{Fetter}, to factorize 
products of operators, of the generic form
$\hat{b}_i^\dagger (\tau)\hat{b}_j(\tau) \hat{a}_k(\tau)$, by
\begin{equation}
\hat{b}_i^\dagger(\tau) \hat{b}_j(\tau) \hat{a}_k (\tau)\approx \langle
\hat{b}_i^\dagger(\tau) \hat{b}_j (\tau)\rangle \hat{a}_k(\tau) - \langle
\hat{b}_i^\dagger(\tau) \hat{a}_k(\tau) \rangle \hat{b}_j(\tau) ,
\end{equation}
where the expectation value is over the state $|\psi_F(0)\rangle$
since we work in the Heisenberg picture. At this level of
approximation, we neglect all contributions from pairing. 
This factorization scheme readily yields the time-dependent
Hartree-Fock equations of motion (in dimensionless variables)
\begin{eqnarray}
\label{equ:finalferm} i\frac{\partial \hat{a}_n}{\partial \tau}
&=& \sum_{k} \left[ \left(A_n \delta_{nk} + Q^{bb}_{nk}\right)
\hat{a}_k - \left({Q^{ab}_{nk}}^* -
\tilde{g}\delta_{nk}\right)\hat{b}_k
\right] \nonumber\\
i\frac{\partial \hat{b}_n}{\partial \tau}
&=& \sum_{k}
\left[
\left(B_n \delta_{nk} + Q^{aa}_{nk}\right) \hat{b}_k -
\left(Q^{ab}_{nk} - \tilde{g}\delta_{nk}\right) \hat{a}_k
\right] \nonumber\\
&+&
C_n \hat{b}_{n+2} + D_n \hat{b}_{n-2},
\end{eqnarray}
where we have introduced the time-dependent coefficients
\begin{eqnarray}
\label{equ:abbreviationnonlinearferm} Q^{aa}_{nk}(\tau) &=&
\tilde{U}_0 \sum_{i,j} U_{nijk}
\langle \hat{a}_i^\dagger(\tau) \hat{a}_j(\tau)\rangle \nonumber\\
Q^{bb}_{nk}(\tau) &=& \tilde{U}_0 \sum_{i,j} U_{nijk}
\langle \hat{b}_i^\dagger (\tau)\hat{b}_j(\tau)\rangle \nonumber\\
Q^{ab}_{nk}(\tau) &=& \tilde{U}_0 \sum_{i,j} U_{nijk} \langle
\hat{a}_i^\dagger(\tau) \hat{b}_j(\tau) \rangle,
\end{eqnarray}
and $\tilde{U}_0=U_0/\hbar\omega_+$ is a dimensionless interaction
strength.

\begin{figure}
\begin{center}
\includegraphics[width=0.9\columnwidth]{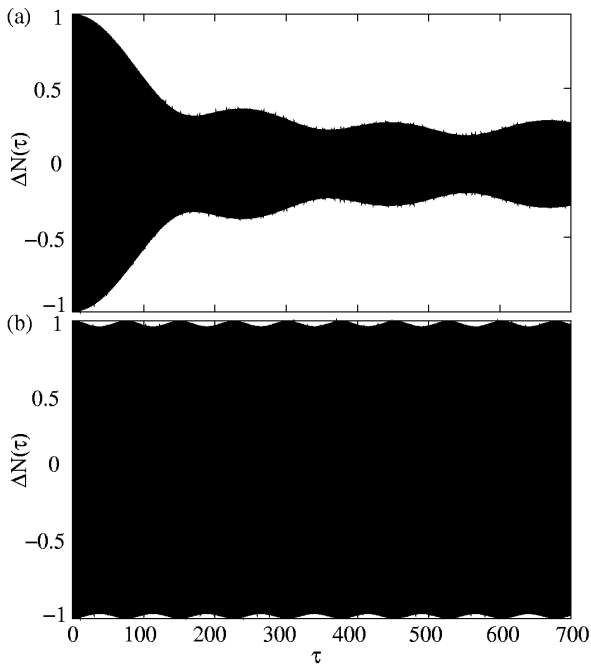}
\vspace{.3cm} \caption{Population difference $\Delta N(\tau)$ for
$N=10$ fermions, a trap ratio
  $\beta=0.9$ and in the strong-coupling regime $\tilde{g}=10.0$.
  The plots, which result from the numerical integration of Eqs.
  (\ref{equ:finalferm}), are for different strengths of the two-body
  collisions: (a) $\tilde{U}_0 = 0.01$; (b) $\tilde{U}_0 = 0.1$.
  Dimensionless time $\tau$ in units of $1/\omega_+$.
\label{fig:figure3} }
\end{center}
\end{figure}

The effect of collisions is illustrated in Figs.
\ref{fig:figure3}a,b, which show the population inversion $\Delta
N(\tau)$ for two values of the interaction strength $\tilde{U}_0$.

For weak enough collisions, the dynamics of the system is not
significantly altered, as should of course be expected. However,
we observe a quantitative change in the dynamics of $\Delta N$ as
$\tilde{U}_0$ is increased. Instead of a collapse and
revivals, $\Delta N(\tau)$ now undergoes nearly full Rabi
oscillations.

\begin{figure}
\begin{center}
\includegraphics[width=0.9\columnwidth]{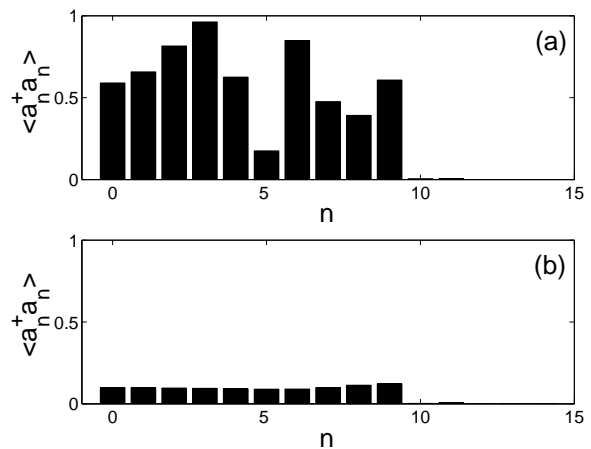}
\vspace{.3cm} \caption{Occupation of the first 15 upper trap
levels at the dimensionless time $\tau = 490/\omega_+$, (a) for
  the parameters of Fig. \ref{fig:figure3}a; and (b) for the
  parameters of Fig. \ref{fig:figure3}b.
\label{fig:figure4} }
\end{center}
\end{figure}

A first hint at the cause of this changed behavior is offered by
Fig. \ref{fig:figure4}, which shows a snapshot of the level
populations in the ``+''-trap for the cases of
Fig. \ref{fig:figure3}a and \ref{fig:figure3}b, respectively. We
observe that the smaller value of $\tilde{U}_0$ corresponds to an
inhomogeneous level population distribution, whereas the higher
nonlinearity causes the trap levels to be almost equally
populated.

\begin{figure}
\begin{center}
\includegraphics[width=0.9\columnwidth]{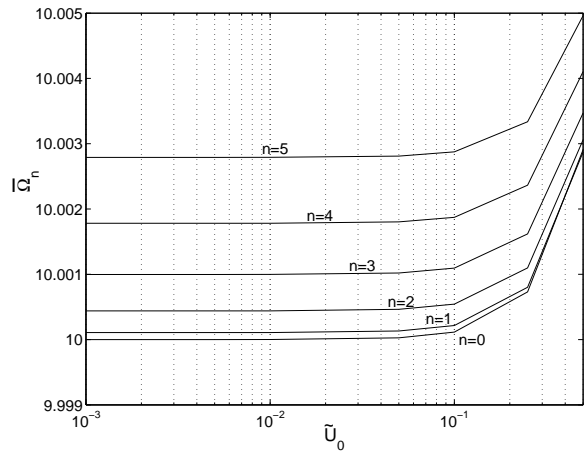}
\vspace{.3 cm} \caption{ Semi-logarithmic plot of the
time-averaged generalized Rabi frequencies for the six lowest trap
levels as a function of the nonlinear parameter, otherwise the
same parameters as in Fig. \ref{fig:figure3} are used. $\bar{\Omega}_n$
is given in units $\omega_+$, $\tilde{U}_0$ in units of $\hbar\omega_+$.
\label{fig:figure5} }
\end{center}
\end{figure}

A more quantitative understanding of the role of collisions can be
gained by estimating how the nonlinear terms in Eq.
(\ref{equ:finalferm}) modify the (collisionless) Rabi frequency. A
numerical evaluation of the coefficients $U_{ijkl}$ shows that
elastic collisions, $i=j=k=l$, dominate the dynamics of the
system. In addition, $U_{nnnn}$ turns out to be a decreasing
function of $n$. Keeping the elastic contributions to the
collision-induced dynamics only, and neglecting as in the
strong-coupling regime of section III the effects of intra-trap
coupling terms $\hat{b}_{n\pm2}$, one find that as a result of
collisions Eq. (\ref{equ:omegan}) is approximately changed to
\begin{equation}
\label{equ:omeganonlinear} \Omega^{NL}_n(\tau) = \sqrt{g^2 +
\frac{1}{4}\left( A_n - B_n + Q^{bb}_{nn}(\tau) -
Q^{aa}_{nn}(\tau) \right)^2}.
\end{equation}
Fig. \ref{fig:figure5} shows, as a function of $\tilde{U}_0$, the
time-dependent Rabi frequencies $\Omega^{NL}_n(\tau)$ averaged
over a time interval $\Theta$ large compared to their inverse,
\begin{equation}
\bar{\Omega}_n = \frac{1}{\Theta}\int_0^\Theta {d}\tau\,
\Omega^{NL}_n(\tau).
\end{equation}
Because $U_{nnnn}$ is a decreasing function of $n$, its
contribution tends to compensate the $n^2$ dependence of Eq.
(\ref{equ:omegan}). As a result, there is a range of collision
strengths for which the dependence of $\Omega^{NL}_n(\tau)$ on $n$
largely disappears. In this range, paradoxically, the dynamics of
the collision-dominated Fermi system resembles that of a
collisionless Bose system. From this admittedly crude argument --
which is however consistent with our full numerical results -- we
also conjecture that for even larger ${\tilde U}_0$, the
approximate cancellation of the $n$-dependence of the Rabi
frequencies will disappear and we expect an overall dephasing and
decay of the population difference $\Delta N(t)$. It has
unfortunately proven prohibitive to try and check this conjecture
numerically.

We now turn to the case of bosonic atoms. Bose statistics allows
for $s$-wave collisions between atoms in the same spin state, so
that the collisional Hamiltonian is now
\begin{eqnarray}
\label{equ:hcolbos} {\hat{\cal H}}_{\rm {col}} &=& U_+ \int {d}x\,
\hat{\Psi}^\dagger_+(x)\hat{\Psi}^\dagger_+(x)
\hat{\Psi}_+(x)\hat{\Psi}_+(x) \nonumber\\
&+& U_-\int {d}x\, \hat{\Psi}^\dagger_-(x)\hat{\Psi}^\dagger_-(x)
\hat{\Psi}_-(x)\hat{\Psi}_-(x)  \nonumber\\
&+& 2 U_x \int {d}x\,
\hat{\Psi}^\dagger_+(x)\hat{\Psi}^\dagger_-(x)
\hat{\Psi}_+(x)\hat{\Psi}_-(x),
\end{eqnarray}
where the $U_i$, $i=\{+,-,x\}$ characterize the strength of the
collisions. In the following we assume for simplicity
$U_+=U_-=U_0$ and $U_x=\eta_xU_0$.

\begin{figure}
\begin{center}
\includegraphics[width=0.9\columnwidth]{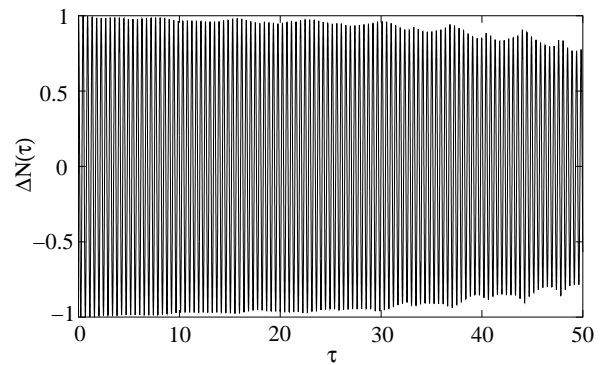}
\vspace{.3cm} \caption{$\Delta N(\tau)$ for $N=20$ bosons, trap
  ratio $\beta=0.9$, linear coupling $\tilde{g}=7.0$ and $\eta_x=1.5$, from
  the numerical solution of Eq. (\ref{equ:finalbos}) with
  $\tilde{U}_0=0.5$.Dimensionless time $\tau$ in units of
  $1/\omega_+$} \label{fig:figure6}.
\end{center}
\end{figure}

\begin{figure}
\begin{center}
\includegraphics[width=0.9\columnwidth]{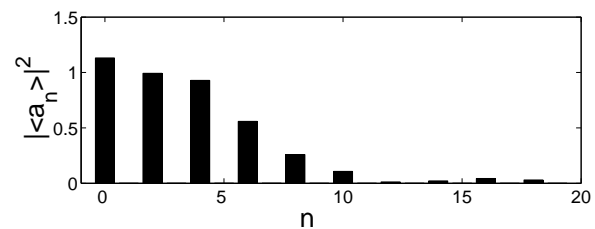}
\vspace{.3cm} \caption{Level population of the ``+''-trap at the
dimensional time $\tau=25/\omega_+$ for the parameters of Fig.
\ref{fig:figure6}. \label{fig:figure7} }
\end{center}
\end{figure}

To truncate the Heisenberg equations of motion for the field
operators, we now invoke a mean-field approximation, factorize all
products of operators, and replace the resulting expectation
values by time-dependent $c$-numbers. This gives
\begin{eqnarray}
\label{equ:finalbos} i\frac{d\langle\hat{a}_n\rangle}{d \tau} &=&
\sum_{k}
    \left(A_n\delta_{nk} + Q^{aa}_{nk} + \eta_x Q^{bb}_{nk}\right)
    \langle\hat{a}_k\rangle + \tilde{g}\langle\hat{b}_n\rangle
    \nonumber\\
i\frac{\partial \langle\hat{b}_n\rangle}{\partial \tau}
&=& \sum_{k}
    \left(B_n\delta_{nk} + Q^{bb}_{nk} + \eta_x Q^{aa}_{nk}\right)
    \langle\hat{b}_k\rangle + \tilde{g}\langle\hat{a}_n\rangle
    \nonumber\\
&+& C_n\langle\hat{b}_{n+2}\rangle + D_n\langle\hat{b}_{n-2}\rangle,
\end{eqnarray}
where
\begin{eqnarray}
\label{equ:abbreviationnonlinearbos} Q^{aa}_{nk}(\tau) &=&
2\tilde{U}_0 \sum_{i,j} U_{n,i,j,k}
\langle \hat{a}_i(\tau)\rangle^*\langle \hat{a}_j (\tau)\rangle \nonumber\\
Q^{bb}_{nk}(\tau) &=& 2\tilde{U}_0 \sum_{i,j} U_{nijk} \langle
\hat{b}_i(\tau)\rangle^*\langle\hat{b}_j (\tau)\rangle,
\end{eqnarray}
and the expectation values are with respect to the state
$|\psi_B(0)\rangle$. Figure \ref{fig:figure6} shows the inversion
$\Delta N(\tau)$ for a sample of bosonic atoms initially in the
internal state $|+\rangle$ for a nonlinear parameter ${\tilde U}_0
= 0.5$. In contrast to the case where collisions are absent and we
have full Rabi oscillations, c.f. Sec. \ref{sec:dynamics}, here
one starts observing a damping of the oscillations. This is
clearly a result of the scattering of atoms into higher trap
states. This is illustrated in Fig. \ref{fig:figure7}, which shows
the population of the first upper trap levels at a fixed time. The
transitions between the populated trapped states are characterized
by $n$-dependent Rabi-frequencies, leading to the onset of a
dephasing process resembling the situation for noninteracting
fermions \cite{footnote1}. We see, then, that in the case of
interacting bosons, intra-trap scattering is an important element
of the dynamics of the Raman coupler, which rapidly evolves to a
multimode behavior; in contrast in the intrinsically multimode
fermionic case $U_{ijkl}$ tends to reduce the spread in Rabi
frequencies and thus inhibits dephasing.

A remarkable property of the bosonic trap population distribution
is that only even trap levels are occupied, see Fig.
\ref{fig:figure7}. This is a combined result of three facts: (a)
at $T=0$ all atoms are initially in the ground state of one of the
traps; (b) $s$-wave scattering only couples trap states of same
parity, as expressed by the symmetry properties of the collision
matrix from Eq. \ref{equ:scatmat}; (c) intra-trap coupling only
couples trap levels with $\Delta n = 2$, see Eq.
(\ref{equ:finalsame}).

It is known from nonlinear optics \cite{Jens82} and atom optics
\cite{Sche95} that systems governed by a pair of coupled nonlinear
Schr{\" o}dinger equations can reach a regime where the nonlinear
phase shifts dominate their dynamics. Such two-mode systems
exhibit Rabi oscillations for small nonlinearities, but
mode-coupling is inhibited above a certain strength of the
nonlinearity. This effect is absent in the present multimode
system, a result of the strong inter-mode scattering.


%===========================================
\section{Summary}
\label{sec:summary}
%===========================================

The Raman coupling between two internal states of a trapped Fermi
gas exhibits a rich dynamics, quite different from its bosonic
counterpart. This is of course due primarily to the fact that a
Fermi gas occupies a large number of trap states, and hence can
never be approximated as a two-mode system. In particular, we have
identified two limiting regimes, dependent upon whether inter-trap
or intra-trap dynamics is dominant. In the general situation where
the traps associated with the two internal states have different
frequencies, the first of these regimes leads to dynamics
characterized by collapses and revivals. However, two-body
collisions can under appropriate conditions inhibit this behavior,
making the collision-dominated Fermi system more similar to a
collisionless Bose system.

The numerical analysis of fermionic systems appears to be
presently limited to very small numbers $N$ of atoms, a result of
the large memory requirements associated with the need to keep
track of a large number of quantum states. Indeed, most of our
numerical results are limited to $N$ on the order of
10 (especially for the collisional calculations), and even
this required rather large computing facilities.
Despite this limitation and its associated lack of quantitative
predictions, our analysis sheds useful light on the dynamics of
trapped Fermi systems, in particular in the presence of
collisions, and will provide useful guidance in understanding more
realistic trapped Fermi gases in three dimensions and with a large
number of fermionic atoms.

\acknowledgements We thank J.~V. Moloney for providing us with CPU
time on his parallel cluster, and E.~M. Wright and C.~P. Search
for valuable discussions. This work is supported in part by the US
Office of Naval Research under Contract No. 14-91-J1205, by the
National Science Foundation under Grant No. PHY-0098129, by the US
Army Research Office, by NASA, and by the Joint Services Optics
Program.


\begin{references}
\bibitem{footnote1}
The choice of a given number of modes in the numerical simulations
limits the evolution time to values such that the population of
modes close to the numerical cut-off point remains small.
\end{references}
\end{document}